\documentclass[article,epsfig,graphicx,nofootinbib,showpacs,eqsecnum,aps,amsmath,mathrsfs,eufrak,amssymb,floatfix,twocolumn]{revtex4}
\usepackage{graphicx,graphics}
\usepackage{epsfig}
\begin{document}

\title{Competing $\gamma$-rigid and $\gamma$-stable vibrations in neutron rich Gd and Dy isotopes\footnote{The final publication is available at \url{http://link.springer.com/article/10.1140/epja/i2015-15126-8}}}
\author{R. Budaca and A. I. Budaca}

\affiliation{Horia Hulubei National Institute of Physics and Nuclear Engineering, RO-077125 Bucharest-Magurele, Romania}
\date{\today}
\begin{abstract}
An exactly separable version of the Bohr Hamiltonian which combines the $\gamma$-stable and $\gamma$-rigid axial vibration-rotation is used to describe the collective properties of few neutron rich transitional nuclei. The coupling between the two types of collective motion is managed through a rigidity parameter which also influences the geometry of the shape-phase space. While the $\gamma$-angular part of the problem associated to axially symmetric shapes is treated within the small angles approximation and the stiff $\gamma$ oscillation hypothesis, the $\beta$ vibration is described by means of a Davidson potential. The resulting model have three free parameters not counting the scale and was successfully applied for the description of the collective spectra for few heavier isotopes of Gd and Dy. In both cases a critical nucleus was identified through a discontinuous behavior in respect to the rigidity parameter and relevant experimental observables.
\end{abstract}
\pacs{21.60.Ev,21.10.Re,27.70.+q}
\maketitle

\section{Introduction}
\label{sec:1}
The rare earth region of the nuclide chart is well known to be the crossroad of various shape phase transitions \cite{Casten}, providing candidate nuclei for all known dynamical symmetries defined in the mainframe of the Interacting Boson Model (IBM) \cite{IBM} and their variations reflected in the Bohr-Mottelson \cite{Bohr1,Bohr2} treatment of the quadrupole collective motion. Higher order multipole symmetries, such as tetrahedral and octahedral ones \cite{Dudek}, are also expected to emerge in this region. As a consequence, most of the numerous collective solutions \cite{Fortunato1,Cejnar} have isolated rare earth experimental realizations, lacking thus the systematization specific to the boson approaches such as IBM \cite{Scholten,Ramos,von,McC} and Coherent State Model \cite{Rabu1,Rabu2,Delion}.

Given the near exhaustion of the exactly and approximatively solvable potentials and the introduction of the Algebraic Collective Model \cite{Rowe1,Rowe2,Rowe3} which allows a rapidly converging diagonalization procedure for solving Bohr Ha\-mil\-tonian for any general potential, alternative approaches are more than welcome. The realization of Bohr-Mottelson model in other than the usual five-dimensional shape phase space is an example of such formalisms. The reformulation of the Bohr Hamiltonian in six dimensions \cite{Georg} explains the origin of the deformation dependent mass term used in \cite{BonMD1,BonMD2} to describe a large variety of nuclei. On the other side, the $\gamma$-rigid regimes lead to simple descriptions of the basic rotation-vibration coupling with quite good agreement with experiment \cite{Davydov,BonZ4,BonX3,Budaca1,Budaca2,Buganu,Buganu2,Zhang}. Going further and speculating the obvious theoretical similarities between the $\gamma$-rigid and $\gamma$-stable hypotheses regarding the quadrupole collective excitations, we proposed in \cite{Noi} a hybrid model based on the interplay between these two already well established approaches. The idea was initially used to combine the $\gamma$-rigid prolate picture with an additional stiff $\gamma$ oscillation around $\gamma=0$ through a $\gamma$-rigidity control parameter. The two kinetic energy operators were considered in connection with an infinite square well potential in $\beta$ shape variable, which lead to a two parameter model with $X(3)$ \cite{BonX3} and ES-$X(5)$ \cite{BonX5ES} as limiting cases. Although the energy spectrum of $X(3)$ was exactly reproduced in the $\gamma$-rigid limit, the wave function properties where found to be altered. In order to remedy this drawback one must investigate the model at its origin. This is done here, by deriving the same $\gamma$-rigid/stable energy through the quantization of a similar classical energy function defined in the framework of the Liquid Drop Model (LDM) \cite{BMB}. Nevertheless, numerical applications of Ref.\cite{Noi} on the energy spectra revealed such an ambiguous behaviour in few neutron rich rare earth nuclei. Although the considered flat $\beta$ potential is suitable for transitional nuclei, it lacks the versatility needed to validate the experimental candidates as being critical in respect to some aspect of collective motion. For this scope, we adopt in this paper the Davidson potential \cite{Davidson} instead of the infinite square well in the $\beta$ variable, which contains an additional parameter but still retains the advantage of being exactly solvable. Indeed, as was shown in Ref.\cite{BonDES}, the exactly separable solution of the Bohr Hamiltonian associated to a Davidson potential in $\beta$ shape variable called ES-D has a large applicability throughout different regions of the nuclide chart. Such that with the resulting model, one can investigate more widely the simultaneous realization of both $\gamma$-rigid and $\gamma$-stable shape phase conditions in the region identified in Ref. \cite{Noi}. As is expected, the model works very well for the $N=96$ isotopes of Gd and Dy treated in the preceding paper as well as for their neighbouring isotopes but with essentially different model characteristics. The peculiarity of the $N=96$ isotopes compared to their neighbours is also discussed in connection to some relevant experimental spectral signatures.

The results are presented as follows. The adaptation to the Davidson potential of the theoretical model introduced in Ref.\cite{Noi} is briefly reviewed in the next Section with a special focus on the $\gamma$-rigid limit. Based on the classical foundations of the model, one establishes the geometry of its shape phase space in Section 3. The model's features and its dependence on the three free parameters is investigated in Section 4. In the next Section are presented the results of the model fits comprising the collective energy spectrum with the associated transition probabilities for three Gd and three Dy isotopes. The final conclusions are drawn in the last Section.

\section{Solution of the $\gamma$-stable/rigid collective Hamiltonian}
\label{sec:2}
Before starting with the analytical foundations of the mo\-del, it is instructive to have a clear understanding what means $\gamma$-rigid, $\gamma$-soft and $\gamma$-stable collective conditions. The difference between the first two was extensively explained in Ref.\cite{Fortunato1} especially due to the confusion surrounding the use of such terminology. Basically, such a characteristic is given by the shape of the potential energy in respect to the $\gamma$ shape variable in this case. If the potential has a very sharp minimum, the corresponding wave functions will be strongly confined in the region of the minimum. The extreme regime of this situation, i.e. when the system can be described by a single value of $\gamma$ is referred to as the $\gamma$-rigidity. From phenomenological point of view, a $\gamma$-rigid system is associated to pure rotations and axial vibrations. In the general case of the collective model, the potential energy minimum is however extended, allowing thus the system to fluctuate around the mean value of the $\gamma$ deformation. As a consequence, the wave functions are also extended, such that the associated shape is then given by a superposition of various $\gamma$ deformations. Such a system is then intuitively called a $\gamma$-soft one. It is worth to mention here that in the $\gamma$-soft models the rotations are no longer separated but coupled to the $\gamma$ oscillations. Finally there are two kinds of $\gamma$-softness, namely the $\gamma$-stable and $\gamma$-unstable. The first is associated to a potential energy which has a localized minimum, while the other one is characterized by the lack of the $\gamma$ potential such that there is no information about the $\gamma$ deformation of the system.

Restricting the $\gamma$ shape variable to certain values one can obtain more simple models \cite{BonZ4,BonX3}. Indeed, choosing $\gamma=0$ one reaches the prolate $\gamma$-rigid version of the collective model whose kinetic energy operator reads \cite{BonX3,Budaca1,Budaca2,Noi}:
\begin{equation}
\hat{T}_{r}=-\frac{\hbar^{2}}{2B}\left[\frac{1}{\beta^{2}}\frac{\partial}{\partial{\beta}}\beta^{2}\frac{\partial}{\partial{\beta}}-\frac{\mathbf{Q}^{2}}{3\beta^{2}}\right],
\label{Tr}
\end{equation}
which describes the basic rotation-vibration coupling. $\mathbf{Q}$ is the angular momentum operator from the intrinsic frame of reference. Allowing nonaxial vibrations through the $\gamma$ shape variable, the above operator recovers the expression from the usual five dimensional Bohr Hamiltonian:
\begin{eqnarray}
\hat{T}_{s}&=&-\frac{\hbar^{2}}{2B}\left[\frac{1}{\beta^{4}}\frac{\partial}{\partial{\beta}}\beta^{4}\frac{\partial}{\partial{\beta}}+\frac{1}{\beta^{2}\sin{3\gamma}}\frac{\partial}{\partial\gamma}\sin{3\gamma}\frac{\partial}{\partial\gamma}\right.\nonumber\\
&&\left.-\frac{1}{4\beta^{2}}\sum_{k=1}^{3}\frac{Q_{k}^{2}}{\sin^{2}{\left(\gamma-\frac{2}{3}\pi k\right)}}\right],
\label{Ts}
\end{eqnarray}
where by $Q_{k}(k=1,2,3)$ are denoted the operators of the total angular momentum projections, while $B$ is the mass parameter. Restricting ourselves in the $\gamma$-soft case only to the collective motion with small oscillations of the $\gamma$ shape variable around the value zero, the rotational term can be very well approximated by \cite{Iachello1}:
\begin{equation}
\sum_{k=1}^{3}\frac{Q_{k}^{2}}{\sin^{2}{\left(\gamma-\frac{2}{3}\pi k\right)}}\approx\frac{4}{3}\mathbf{Q}^{2}+Q_{3}^{2}\left(\frac{1}{\sin^{2}{\gamma}}-\frac{4}{3}\right).
\label{app}
\end{equation}

In a preceding paper \cite{Noi} the description of a combined $\gamma$-rigid and $\gamma$-soft nucleus was approached by considering the Hamiltonian:
\begin{equation}
H=\chi \hat{T}_{r}+(1-\chi)\hat{T}_{s}+V(\beta,\gamma),
\label{Ht}
\end{equation}
where $0\leq\chi<1$ measures the system's $\gamma$-rigidity. The $\gamma$-rigid limit $\chi=1$ was avoided in order to preserve the five dimensional geometry of the curvilinear space, because the resulting Hamiltonian have five degrees of freedom. The exact separation of the $\beta$ variable from the $\gamma$-angular ones is achieved by considering the following expression for the potential energy:
\begin{equation}
v(\beta,\gamma)=\frac{2B}{\hbar^{2}}V(\beta,\gamma)=u(\beta)+(1-\chi)\frac{u(\gamma)}{\beta^{2}},
\label{v}
\end{equation}
a form similar to that used in Refs. \cite{Fortunato1,BonX5ES,Wilets,Fortunato2,Fortunato3}. Factorizing the total wave function as $\Psi(\beta,\gamma,\Omega)=\xi(\beta)\eta(\gamma)$ $\times D^{L}_{MK}(\Omega)$ where $D^{L}_{MK}$ are the Wigner functions with $\Omega$ denoting the set of three Euler angles and $L$ being the total angular momentum, while $M$ and $K$ - its projections on the body-fixed and respectively laboratory-fixed z axis, the associated Schr\"{o}dinger equation is separated into a $\beta$ part:
\begin{equation}
\left[-\frac{\partial^{2}}{\partial{\beta^{2}}}-\frac{2(2-\chi)}{\beta}\frac{\partial}{\partial{\beta}}+\frac{W}{\beta^{2}}+u(\beta)\right]\xi(\beta)=\epsilon\xi(\beta),
\label{b}
\end{equation}
where $\epsilon=\frac{2B}{\hbar^{2}}E$, and a $\gamma$-angular one. Averaging the last on the Wigner states and using the approximation (\ref{app}) one obtains the following equation for the $\gamma$ shape variable:
\begin{equation}
\left[-\frac{1}{\sin{3\gamma}}\frac{\partial}{\partial\gamma}\sin{3\gamma}\frac{\partial}{\partial\gamma}+\frac{K^{2}}{4\sin^{2}{\gamma}}+u(\gamma)\right]\eta(\gamma)=\epsilon_{\gamma}\eta(\gamma),
\label{ecg}
\end{equation}
with
\begin{equation}
\epsilon_{\gamma}=\frac{1}{1-\chi}\left[W-\frac{L(L+1)-(1-\chi)K^{2}}{3}\right].
\label{egw}
\end{equation}
Applying a harmonic approximation for the trigonometric functions around $\gamma=0$ and adopting a harmonic oscillator form for the $\gamma$ potential
\begin{equation}
u(\gamma)=\left(3a\right)^2\frac{\gamma^{2}}{2},
\label{ugamma}
\end{equation}
as in Refs.\cite{BonX5ES,Iachello1}, leads to the following differential equation for $\gamma$ shape variable:
\begin{equation}
\left[-\frac{1}{\gamma}\frac{\partial}{\partial\gamma}\gamma\frac{\partial}{\partial\gamma}+\left(\frac{K}{2}\right)^{2}\frac{1}{\gamma^{2}}+(3a)^{2}\frac{\gamma^{2}}{2}\right]\eta(\gamma)=\epsilon_{\gamma}\eta(\gamma).
\label{ecga}
\end{equation}
It resembles the radial equation for a two dimensional harmonic oscillator, with the solutions given in terms of the Laguerre polynomials \cite{Iachello1}:
\begin{equation}
\eta_{n_{\gamma}K}(\gamma)=N_{nK}\gamma^{\left|\frac{K}{2}\right|}e^{-3a\frac{\gamma^{2}}{2}}L_{n}^{|\frac{K}{2}|}(3a\gamma^{2}),
\label{fg}
\end{equation}
where $N_{nK}$ is a normalization constant, $n=(n_{\gamma}-|K|/2)/2$ and $a$ is a parameter associated through the string constant of the harmonic oscillator potential (\ref{ugamma}) to the stiffness of $\gamma$ vibrations. The corresponding eigenvalues are
\begin{equation}
\epsilon_{\gamma}=3a(n_{\gamma}+1),\,\,n_{\gamma}=0,1,2,...,
\label{eg}
\end{equation}
with $K=0,\pm 2n_{\gamma}$ for $n_{\gamma}$ even and $K=\pm 2n_{\gamma}$ for $n_{\gamma}$ odd, respectively. Note also that for $K=0$ the rotational sequence is of the form $L=0,2,4,..$ specific to ground and $\beta$ bands, while for $K>0$ associated to nonzero $\gamma$ vibration quanta, it is described by the $L=K,K+1,K+2,..$ rule. This treatment of the $\gamma$ degree of freedom is combined with the use of the Davidson potential \cite{Davidson}
\begin{equation}
u(\beta)=\beta^{2}+\frac{\beta_{0}^{4}}{\beta^{2}},
\end{equation}
in the $\beta$ equation (\ref{b}). $\beta_{0}$ represents the minimum of the potential, such that when its vanishing one obtains the exactly solvable harmonic oscillator model $X(5)$-$\beta^{2}$ \cite{BonX5ES}, while for $\beta_{0}\rightarrow\infty$ the model tends to the $SU(3)$ limit. Such a choice for the potentials $u(\beta)$ and $u(\gamma)$ is suitable for axially symmetric prolate nuclei which represent the majority of quadrupole deformed isotopes, and was successfully used to describe the collective spectra for a large number of nuclei ranging from lanthanides to actinides \cite{BonDES}. The $\beta$ eigenvalue problem for the Davidson potential in the usual Bohr-Mottelson model is exactly solvable \cite{BonDES,RoweD}, such that the solution of equation (\ref{b}) is readily transposed as:
\begin{equation}
\xi_{LKn_{\beta}n_{\gamma}}(\beta)=N_{n_{\beta}p}(\chi)\beta^{p+\chi}e^{-\frac{\beta^{2}}{2}}L_{n}^{p+\frac{3}{2}}(\beta^{2}),
\label{fb}
\end{equation}
where $L_{n}^{k}$ are the associated Laguerre polynomials and $N_{nL}(\chi)$ is the normalization constant, while $p$ is defined as:
\begin{eqnarray}
p&=&-\frac{3}{2}+\left[\frac{L(L+1)-(1-\chi)K^{2}}{3}+\beta_{0}^{4}+\right.\nonumber\\
&&\left.\left(\frac{3}{2}-\chi\right)^{2}+(1-\chi)3a(n_{\gamma}+1)\right]^{\frac{1}{2}}.
\end{eqnarray}
With the corresponding eigenvalue being
\begin{eqnarray}
\epsilon_{LKn_{\beta}n_{\gamma}}&=&2n_{\beta}+1+\nonumber\\
&&\left[\frac{L(L+1)-(1-\chi)K^{2}}{3}+\beta_{0}^{4}+\right.\nonumber\\
&&\left.\left(\frac{3}{2}-\chi\right)^{2}+(1-\chi)3a(n_{\gamma}+1)\right]^{\frac{1}{2}},
\label{vare}
\end{eqnarray}
the total energy of the system normalized to the ground state is finally expressed as:
\begin{equation}
E_{LKn_{\beta}n_{\gamma}}=\frac{\hbar^{2}}{2B}\left[\epsilon_{LKn_{\beta}n_{\gamma}}-\epsilon_{0000}\right].
\label{Et}
\end{equation}
A closer look at Eq.(\ref{vare}) shows that the ground state becomes infinitely degenerate in respect with $\gamma$ vibrational quanta $n_{\gamma}=$ even when $\chi=1$, fact which contravenes with the finite structure of the atomic nucleus. However this is not actually true because the scale of the $\gamma$ contribution to the total energy is a product of stiffness and rigidity and both of them are interdependent. Indeed, the $\gamma$-rigid limit also means an infinite stiffness of the $\gamma$ oscillations, such that the corresponding contribution is actually indefinite, spoiling thus the infinite degeneracy of the ground state. This fact is then consistent with the $\gamma$-rigid models where there are no $\gamma$ excited states. In order to show the relation between the stiffness and the rigidity parameters, it is useful to introduce a renormalized parameter $c=(1-\chi)a$. In this way the $\gamma$ energy (\ref{eg}) has a clearly understood behaviour as function of $\chi$, becoming infinite when $\chi=1$ which actually means $\gamma$-rigidity. This result is evident also from the alternative representation (\ref{egw}) of the $\gamma$ energy without appealing to the mentioned renormalization of the stiffness parameter. In the same way changes the scale of the $\gamma$ function (\ref{fg}), such that the associated probability distribution for $\chi\rightarrow 1$ acquires a strongly localized maximum at a value which tends asymptotically to zero. In conclusion, the $a$ representation gives a consistent description of the Hamiltonian and a correct energy spectrum but has a poor description of the states in the $\gamma$-rigid case. In contradistinction, the use of $c$ parametrization has a complementary role. However, the case $\chi=1$ must be treated separately and as it has only a theoretical importance, in the numerical applications one will stick to the $a$ parametrization.

\section{The model's shape phase space}
\label{sec:3}
In what follows one will show that an identical differential equation for determining the energy of the system is obtained if one starts from the classical picture of LDM \cite{BMB}, clarifying thus the mathematical foundations of the approach introduced in Ref.\cite{Noi}. Moreover, as will be shown, such an alternative picture allows a consistent description of the $\gamma$-rigid limit. For this, one recalls that the LDM classical kinetic energy is quadratic in the time derivatives of all variables and it can be separated into vibrational and rotational parts:
\begin{eqnarray}
T_{vibr}&=&\frac{B}{2}\left(\dot{\beta}^{2}+\beta^{2}\dot{\gamma}^{2}\right),\\
T_{rot}&=&\frac{1}{2}\sum_{k=1}^{3}\omega_{k}^{2}\mathcal{I}_{k}=\sum_{k=1}^{3}\frac{\hbar^{2}I_{k}^{2}}{2\mathcal{I}_{k}},
\end{eqnarray}
with $\mathcal{I}_{k}=4B\beta\sin^{2}{\left(\gamma-2\pi k/3\right)}$ being the moments of inertia while $\omega_{k}$ are the angular frequencies associated to the principal axes indexed by $k$ having as canonical conjugates the classical components of angular momentum $I_{k}$. Considering a similar weighting combination as in (\ref{Ht}), one obtains the following classical function:
\begin{eqnarray}
\mathcal{H}&=&\frac{B}{2}\dot{\beta}^{2}+(1-\chi)\frac{B}{2}\beta^{2}\dot{\gamma}^{2}+\nonumber\\
&&(1-\chi)T_{rot}^{\gamma\neq0}+\chi T_{rot}^{\gamma=0}+V(\beta,\gamma).
\label{Hc}
\end{eqnarray}
The quantization procedure is in general a very complex endeavour, especially in the case of mixed terms of coordinate and conjugate momenta \cite{Rot}. Due to its consistent geometrical construction, LDM has a well established recipe for obtaining a quantum Hamiltonian. Indeed, its kinetic energy operator is given by a Laplacian in a generalized coordinate system \cite{RoweB}:
\begin{equation}
\hat{T}=-\frac{\hbar^{2}}{2}\nabla^{2}=-\frac{\hbar^{2}}{2}\sum_{lm}\frac{1}{J}\frac{\partial}{\partial{x^{l}}}J\bar{g}^{lm}\frac{\partial}{\partial{x^{m}}},
\label{Pauli}
\end{equation}
where $J=\sqrt{det(g)}$ is the Jacobian of the transformation from the quadrupole coordinates $\{q_{k}\}$ to the curvilinear ones $\{x^{l}\}$ defined by the metric tensor:
\begin{equation}
g_{lm}=\sum_{k}\frac{\partial{q_{k}}}{\partial{x^{l}}}\frac{\partial{q_{k}}}{\partial{x^{m}}},\,\,\,\bar{g}^{lm}=\sum_{k}\frac{\partial{x^{k}}}{\partial{q_{l}}}\frac{\partial{x^{k}}}{\partial{q_{m}}}.
\end{equation}
In the Bohr model, the generalized curvilinear coordinates are the two shapes variables $\beta$ and $\gamma$, and the three Euler angles, such that $J=2\beta^{4}|\sin{3\gamma}|$ while the matrix elements of the metric tensor are calculated by using the expression of the quadrupole coordinates as function of $(\beta,\gamma,\Omega)$:
\begin{eqnarray}
q_{m}(\beta,\gamma,\Omega)&=&\beta\Big\{D^{2}_{m0}(\Omega)\cos{\gamma}+\nonumber\\
&&\frac{1}{\sqrt{2}}\left[D_{m2}^{2}(\Omega)+D_{m-2}^{2}(\Omega)\right]\sin{\gamma}\Big\}.
\end{eqnarray}
Finally, applying the formula (\ref{Pauli}) one obtains the well known expression of the kinetic term in the Bohr-Mottel\-son Hamiltonian which is just the Laplacian in the new curvilinear coordinates. This is the consequence of the fact that the classical kinetic energy of the Bohr model can be written as the time derivative of the squared line element $(ds)^{2}=\sum_{k}dq^{k}dq_{k}$. Unfortunately this is not the case of the classical function (\ref{Hc}) which however spans the same five dimensional space with the same metric. In order to quantize the classical Hamiltonian (\ref{Hc}) one must appeal to a modified quantization rule (\ref{Pauli}). Such a prescription was given in Ref.\cite{Proc} in connection to generalized collective systems which do not posses a classical counterpart in terms of the LDM geometry, and reads
\begin{equation}
\hat{T}=-\frac{\hbar^{2}}{2}\sum_{lm}\frac{1}{J}\frac{\partial}{\partial{x^{l}}}J\bar{G}^{lm}\frac{\partial}{\partial{x^{m}}},
\end{equation}
where $G_{lm}$ is a symmetric positive-definite bitensor matrix which is not necessary related to the metric tensor $g_{lm}$. As the classical Hamiltonian (\ref{Hc}) cannot be constructed by considering more involved expressions for the LDM inertial functions because these are restricted by symmetry obeying conditions, its associated quantum Hamiltonian must be constructed by means of the alternative method. Incorporating the $\chi$ dependence in $\bar{G}^{lm}$ and using the above quantization recipe the quantum Hamiltonian associated to the classical function $\mathcal{H}$ is then expressed as:
\begin{eqnarray}
&&H=-\frac{\hbar^{2}}{2B}\left[\frac{1}{\beta^{4}}\frac{\partial}{\partial{\beta}}\beta^{4}\frac{\partial}{\partial{\beta}}+\frac{1-\chi}{\beta^{2}\sin{3\gamma}}\frac{\partial}{\partial\gamma}\sin{3\gamma}\frac{\partial}{\partial\gamma}\right.\nonumber\\
&&\left.-\frac{1-\chi}{4\beta^{2}}\sum_{k=1}^{3}\frac{Q_{k}^{2}}{\sin^{2}{\left(\gamma-\frac{2}{3}\pi k\right)}}+\chi\frac{\mathbf{Q}^{2}}{3}\right]+V(\beta,\gamma),
\end{eqnarray}
where the notations of the precedent section are used. Following the same steps as before, the only difference is met in the $\beta$ equation which reads:
\begin{equation}
\left[-\frac{\partial^{2}}{\partial{\beta^{2}}}-\frac{4}{\beta}\frac{\partial}{\partial{\beta}}+\frac{W}{\beta^{2}}+u(\beta)\right]\xi(\beta)=\epsilon\xi(\beta).
\label{bc}
\end{equation}
Indeed, the coefficients of the first derivative in (\ref{b}) and (\ref{bc}) are different. As a consequence, the solution of the above equation will be $\xi'(\beta)=\beta^{-\chi}\xi(\beta)$. Note, that the present solution was obtained within a space characterized by $J=2\beta^{4}|\sin{3\gamma}|$, whose volume element and respectively the integration measure is proportional to $J$. As the integration metric associated with the function $\xi'(\beta)$ is $\beta^{4}d\beta$, in case of the function (\ref{fb}) the integration measure must be $\beta^{4-2\chi}d\beta$ in order to have both model derivations equivalent. In this way not only the energy but also the wave functions will have a consistent behaviour in the $\gamma$-rigid limit. The total solution of the Hamiltonian (\ref{Ht}), is then given by the normalized product of angular, $\beta$ and $\gamma$ wave functions:
\begin{eqnarray}
&&\Psi_{LMKn_{\beta}n_{\gamma}}(\beta,\gamma,\Omega)=\xi_{LKn_{\beta}n_{\gamma}}(\beta)\eta_{n_{\gamma}|K|}(\gamma)\nonumber\\
&&\times\sqrt{\frac{2L+1}{16\pi^{2}(1+\delta_{K,0})}}\left[D_{MK}^{L}(\Omega)+(-)^{L}D_{M-K}^{L}(\Omega)\right]\nonumber\\
\label{ft}
\end{eqnarray}
which is also symmetrized against time reversal operation \cite{BonX5ES,Iachello1}. In virtue of the above analysis, its normalization is defined within the integration measure $\beta^{4-2\chi}|\sin{3\gamma}|\times$ $d\beta d\gamma d\Omega$ instead of $\beta^{4}|\sin{3\gamma}|d\beta d\gamma d\Omega$ as in Ref.\cite{Noi}, such that the norm of the $\beta$ state (\ref{fb}) acquires the following expression:
\begin{equation}
N_{n_{\beta}p}(\chi)=\sqrt{\frac{2n_{\beta}!}{\left(n_{\beta}+p+\frac{5}{2}\right)!}}.
\end{equation}

\begin{figure*}[th!]
\begin{center}
\includegraphics[clip,trim = 0mm 0mm 0mm 0mm,width=0.98\textwidth]{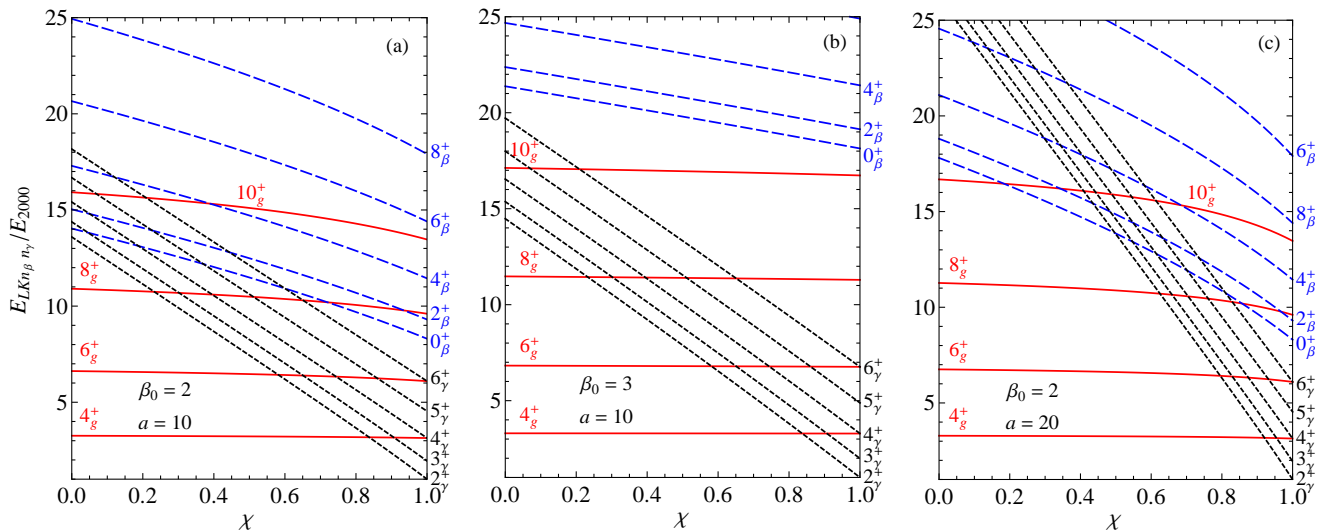}
\end{center}
\vspace{-0.2cm}
\caption{The low lying energy spectrum given as function of the rigidity parameter $\chi$ for: $\beta_{0}=2,a=10$ (a), $\beta_{0}=3,a=10$ (b) and $\beta_{0}=2,a=20$ (c). The ground band energy curves are visualized as solid lines, those corresponding to $\beta$ band by dashed lines, while the dotted lines are associated to the $\gamma$ band states.}
\label{spec}
\end{figure*}

Using the prescriptions of Refs.\cite{BonDES,Bijker} it is straightforward to calculate $E2$ transition rates with the wave functions (\ref{ft}). Basically, using the quadrupole transition operator,
\begin{equation}
T_{m}^{(E2)}=tq_{m},
\label{TE2}
\end{equation}
where $t$ is a scaling factor, one can write the $E2$ transition probability in the factorized form \cite{BonDES,Bijker}:
\begin{eqnarray}
B(E2,LKn_{\beta}n_{\gamma}\rightarrow L'K'n'_{\beta}n'_{\gamma})=\nonumber\\
\frac{5t^{2}}{16\pi}\left(C^{L\,\,2\,\,L'}_{KK'-KK'}B_{L'K'n'_{\beta}n'_{\gamma}}^{LKn_{\beta}n_{\gamma}}G_{K'n'_{\gamma}}^{Kn_{\gamma}}\right)^{2},
\end{eqnarray}
where $B$ and $G$ are integrals corresponding to the shape variables $\beta$ and $\gamma$ with the integration measures $\beta^{4-2\chi}d\beta$ and respectively $|\sin{3\gamma}|d\gamma$, while $C$ is the Clebsch-Gordan coefficient resulting from the angular scalar product. The $\Delta K=0$ transitions being described only by the first term of $E2$ operator (\ref{TE2}), its corresponding $\gamma$ integral reduces to the orthogonality condition for $\eta_{n_{\gamma},|K|}(\gamma)$ wave functions. Similarly in the rigid rotor case, the $B$ and $G$ factors are dropped, such that its transitions are defined solely by the angular momentum selection rules of the Clebsch-Gordan coefficient.

\section{Numerical results}
\label{sec:4}

The model presented in the last section has three free parameters excepting the scale, namely the stiffness of the $\gamma$ oscillations $a$, the minimum of the Davidson potential $\beta_{0}$ and the control parameter $\chi$ which measures the degree of the system's $\gamma$-rigidity. The effect of the three parameters on the energy spectrum can be easily inferred from (\ref{Et}). Moreover, the ground and $\beta$ bands energies actually depend on a single quantity which gathers all three parameters
\begin{equation}
\beta_{0}^{4}+\left(\frac{3}{2}-\chi\right)^{2}+(1-\chi)3a.
\end{equation}
However, when considering also the $\gamma$ band, the three parameters become active and with a well defined individual significance and contribution. The dependence of the low lying energy states on $\chi$ is given in Fig.\ref{spec} for different values of the remaining two parameters expected in the actual applications.

From Fig.\ref{spec} one can see that the ground band states are not very much influenced by $\chi$. Only for higher angular momentum states a small suppression is observed in the high $\chi$ region. This feature is amplified for higher $a$ values and hindered when $\beta_{0}$ is increased. As a matter of fact for $a\rightarrow\infty$ the model's ground band tends to the $SU(3)$ rotational behaviour in the lower $\chi$ part while its upper limit at $\chi\rightarrow1$ is not affected at all. The first observation is easily understood by recalling that an increased stiffness means a more sharpened and confined total $\beta$ potential consistent with the $\beta$-rigid behaviour \cite{Caprio}. Similarly the $a$ independence of the upper limit comes out from the $\gamma$-rigidity hypothesis of the model. In contradistinction, increasing $\beta_{0}$ both extremes of the ground band energy curves are raised asymptotically to the rotational limit. The reason for this similar behavior of the ground band with different intensity at low $\chi$ values as function of $a$ and $\beta_{0}$ is that the $\beta_{0}$ term contribution have the same role as the $\gamma$ energy but with different orders for the involved parameters. Also while the $\gamma$ oscillation ceases in the other limit, the evolution of the remaining $\beta$ potential as function of $\beta_{0}$ explained in the last section shifts the model to conditions appropriate for more deformed nuclei.

The behavior of the $\beta$ bands shown in Fig.\ref{spec} is basically the same as in the ground band case, however without the saturation feature present in the last. This is actually transformed into a linearity effect which is maintained close to the $\gamma$-rigid limit. Also as the $\beta_{0}$ value increases the corresponding energy curves lose in $\chi$ dependence. This is actually the consequence of closeness to the $\beta$-rigid compatible conditions. The effect of the $a$ increase is quite opposite. However in both cases (Fig.\ref{spec}(b) and (c)) the $\beta$ spectrum altogether is shifted to higher energies.

In what concerns the $\gamma$ band energy curves, these are turned out as straight lines whose slopes increase with both $a$ and $\beta_{0}$. The variation of the last induces a much slower growth in the slope but increases the energy distance between two consecutive levels.

As the spectral signatures such as the ratio between the first two excited states, the $\beta$ and $\gamma$ band heads normalized to the energy of the first excited state:
\begin{eqnarray}
R_{4/2}=\frac{E_{4000}}{E_{2000}},\,R_{0/2}=\frac{E_{0010}}{E_{2000}},\,R_{2\gamma/2}=\frac{E_{2201}}{E_{2000}},
\end{eqnarray}
are often used to define the applicability range of a certain model, an analysis in this sense is necessarily required at least for few emerging limiting cases. The realisation of the ES-D and ES-$X(5)$-$\beta^{2}$ when $\chi=0$ in $\beta_{0}>0$ and $\beta_{0}=0$ cases is obvious. In the other limit one obtains the $X(3)$-D model never explored before which comes down to the $X(3)$-$\beta^{2}$ solution \cite{Budaca1} for $\beta_{0}=0$. Although the $\gamma$-rigid model $X(3)$-$\beta^{2}$ does not allow a $\gamma$ band by its construction, in the present model there is no such a restriction when $\chi\rightarrow1$, giving thus the corresponding $\gamma$ band head limit $R_{2\gamma/2}=1$ regardless of the $\beta_{0}$ value. However, the case of $\chi=1$ must be treated separately, keeping in mind the analysis made in Section 2.

\begin{figure}[th!]
\begin{center}
\includegraphics[clip,trim = 0mm 0mm 0mm 0mm,width=0.48\textwidth]{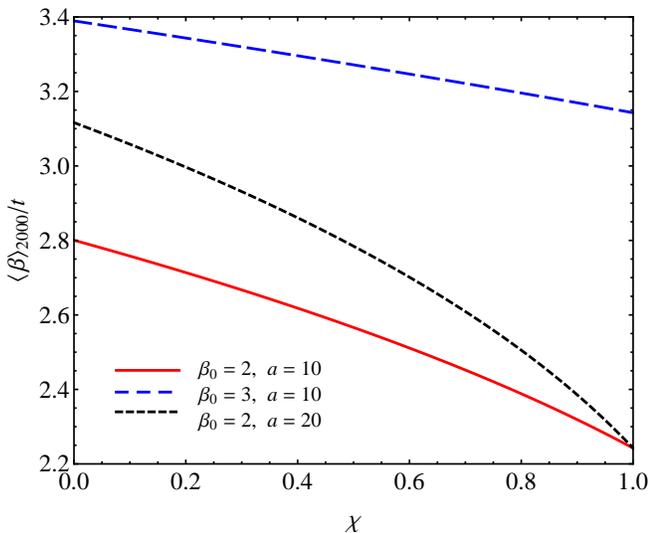}
\end{center}
\vspace{-0.2cm}
\caption{The average $\beta$ deformation in the first excited state given in terms of the scaling constant $t$ as function of rigidity $\chi$ for three $(\beta_{0},a)$ sets of values.}
\label{beta}
\end{figure}

Apart from the spectral characteristics, it is also interesting to see what are the model predictions regarding the deformation of nuclei with such ambiguous collective features. For this, one plotted in Fig.\ref{beta} the mean value of axial deformation $\beta$ in the first excited state as function of the rigidity $\chi$ for the same set of values for parameters $\beta_{0}$ and $a$ as in Fig.\ref{spec}. This average is proportional to the quadrupole momentum which is the experimentally attainable observable. The quantity from Fig.\ref{beta} is given in terms of the scaling parameter $t$ which is arbitrary, such that a direct comparison with the absolute values of the quadrupole moments and electromagnetic transition is irrelevant. However, Fig.\ref{beta} serves for asserting the evolution of the mean deformation along the rigidity variation. The quantities depicted in Fig.\ref{beta} behave as the $\beta$ excited bands from Fig.\ref{spec}. Also from this representation one can see the effect of the $\gamma$ oscillations on the mean deformation, which moves its value toward the outer wall of the total potential (\ref{v}). Consequently, the smallest deformation is achieved in the $\gamma$-rigid limit where the $\gamma$ contribution to the potential and energy vanishes. The obtained mean deformations are in the same domain of values as in other approaches with Davidson potential \cite{BonDES,BonMD1}. It is worth to mention here that the present $\beta$ values are scaled versions of the quadrupole deformation values used for example in microscopic formalisms. The scaling comes from the convention used in Bohr-Mottelson approaches regarding the simplification of the reduced potential by dropping all scaling factors.

\section{Experimental realization}
\label{sec:5}
The model was applied to nuclei placed around $^{160}$Gd, $^{162}$Dy and $^{166}$Er treated previously in Ref.\cite{Noi} with an infinite square well $\beta$ potential. Only for the first two nuclei one obtained similar or better agreement with experiment. Moreover, as the proposed model was expected to be more flexible in reproduction of experimental data than the preceding approach \cite{Noi}, it is also realized in both lighter and heavier neighboring isotopes of $^{160}$Gd and $^{162}$Dy. The selection of model's candidate nuclei was achieved by fitting their experimental energy spectrum comprising ground, $\gamma$ and $\beta$ bands with the energy formula (\ref{Et}), both being normalized to the corresponding energy of the first excited state. The mentioned nuclei were thus found to have the smallest deviations from the experimental data, judging by the quantity
\begin{equation}
\sigma=\sqrt{\frac{1}{N-1}\sum_{i=1}^{N}\left[\frac{E_{i}(Th)}{E_{2_{g}^{+}}(Th)}-\frac{E_{i}(Exp)}{E_{2_{g}^{+}}(Exp)}\right]^{2}},
\label{sigma}
\end{equation}
where $E_{i}(Th)$ is the value calculated using Eq.(\ref{Et}) and $i$ goes over the states of all considered bands.

\setlength{\tabcolsep}{6.5pt}
\begin{table}[th!]
\caption{Theoretical results for ground, $\beta$ and $\gamma$ bands energies normalized to the energy of the first excited state $2_{g}^{+}$ are compared with the available experimental data for $^{158}$Gd\cite{158Gd}, $^{160}$Gd\cite{160GdDy} and $^{162}$Gd\cite{162GdDy}. The dimensionless parameters $\chi,a$ and $\beta_{0}$ are also given together with the corresponding deviation $\sigma$ defined by (\ref{sigma}).}
\label{tab:1}
\begin{center}
\begin{tabular}{ccccccc}
\hline\noalign{\smallskip}
&\multicolumn{2}{c}{$^{158}$Gd}&\multicolumn{2}{c}{$^{160}$Gd}&\multicolumn{2}{c}{$^{162}$Gd}\\
\noalign{\smallskip}\hline\noalign{\smallskip}
$L$&Exp.&Th.&Exp.&Th.&Exp.&Th.\\
\noalign{\smallskip}\hline\noalign{\smallskip}
 $2_{g}^{+}$&   1.00 & 1.00 & 1.00 &  1.00 &  1.00 & 1.00\\
 $4_{g}^{+}$&   3.29 & 3.27 & 3.30 &  3.29 &  3.30 & 3.30\\
 $6_{g}^{+}$&   6.78 & 6.66 & 6.84 &  6.79 &  6.84 & 6.81\\
 $8_{g}^{+}$&  11.37 &11.00 &11.53 & 11.37 & 11.54 &11.43\\
 $10_{g}^{+}$& 16.98 &16.11 &17.28 & 16.90 & 17.29 &17.03\\
 $12_{g}^{+}$& 23.47 &21.84 &24.00 & 23.24 & 24.00 &23.48\\
 $14_{g}^{+}$&       &28.06 &31.59 & 30.26 & 31.57 &30.66\\
 $16_{g}^{+}$&       &34.67 &39.97 & 37.87 & 39.90 &38.45\\
 $18_{g}^{+}$&       &41.58 &      & 45.95 &       &46.77\\
 $20_{g}^{+}$&       &48.74 &      & 54.44 &       &55.53\\
\noalign{\smallskip}\hline\noalign{\smallskip}
 $0_{\beta}^{+}$& 15.04 & 14.80  &  18.33  & 19.34 &19.93&20.46\\
 $2_{\beta}^{+}$& 15.84 & 15.80  &  19.08  & 20.34 &20.84&21.46\\
 $4_{\beta}^{+}$& 17.69 & 18.06  &         & 22.63 &     &23.76\\
 $6_{\beta}^{+}$& 20.58 & 21.45  &         & 26.13 &     &27.27\\
 $8_{\beta}^{+}$&       & 25.79  &         & 30.71 &     &31.89\\
 $10_{\beta}^{+}$&      & 30.91  &         & 36.24 &     &37.49\\
\noalign{\smallskip}\hline\noalign{\smallskip}
 $2_{\gamma}^{+}$& 14.93& 15.34  &13.13& 13.47 &12.07&12.08\\
 $3_{\gamma}^{+}$& 15.92& 16.12  &14.05& 14.34 &12.99&12.98\\
 $4_{\gamma}^{+}$& 17.08& 17.15  &15.25& 15.50 &14.18&14.17\\
 $5_{\gamma}^{+}$& 18.63& 18.42  &16.76& 16.92 &     &15.63\\
 $6_{\gamma}^{+}$& 20.42& 19.91  &18.51& 18.61 &     &17.37\\
 $7_{\gamma}^{+}$&      & 21.61  &20.58& 20.55 &     &19.36\\
 $8_{\gamma}^{+}$&      & 23.52  &22.81& 22.72 &     &21.60\\
 $9_{\gamma}^{+}$&      & 25.60  &     & 25.12 &     &24.07\\
 $10_{\gamma}^{+}$&     & 27.85  &28.14& 27.73 &     &26.76\\
 $11_{\gamma}^{+}$&     & 30.26  &     & 30.54 &     &29.66\\
 $12_{\gamma}^{+}$&     & 32.81  &34.31& 33.53 &     &32.76\\
 $13_{\gamma}^{+}$&     & 35.50  &     & 36.69 &     &36.03\\
 $14_{\gamma}^{+}$&     & 38.30  &     & 40.02 &     &39.47\\
 \noalign{\smallskip}\hline\noalign{\smallskip}
$\chi$&\multicolumn{2}{c}{3$\cdot10^{-4}$} &\multicolumn{2}{c}{0.826} &\multicolumn{2}{c}{0.092}\\
$a$  &\multicolumn{2}{c}{11.349}&\multicolumn{2}{c}{51.538}&\multicolumn{2}{c}{9.052}\\
$\beta_{0}$  &\multicolumn{2}{c}{2.044}&\multicolumn{2}{c}{2.840}&\multicolumn{2}{c}{2.963}\\
$\sigma$  &\multicolumn{2}{c}{0.601}&\multicolumn{2}{c}{0.768}&\multicolumn{2}{c}{0.574}\\
\noalign{\smallskip}\hline
\end{tabular}
\end{center}
\end{table}

\begin{table}[ht!]
\caption{Same as in Table \ref{tab:1} but for $^{160}$Dy\cite{160GdDy}, $^{162}$Dy\cite{162GdDy} and $^{164}$Dy\cite{164Dy}. The value in parentheses denote a state with uncertain band assignment and therefore was not taken into account in the fitting procedure.}
\label{tab:2}
\begin{center}
\begin{tabular}{ccccccc}
\hline\noalign{\smallskip}
&\multicolumn{2}{c}{$^{160}$Dy}&\multicolumn{2}{c}{$^{162}$Dy}&\multicolumn{2}{c}{$^{164}$Dy}\\
\noalign{\smallskip}\hline\noalign{\smallskip}
$L$&Exp.&Th.&Exp.&Th.&Exp.&Th.\\
\noalign{\smallskip}\hline\noalign{\smallskip}
 $2_{g}^{+}$&   1.00 & 1.00 & 1.00 &  1.00 &  1.00 & 1.00\\
 $4_{g}^{+}$&   3.27 & 3.27 & 3.29 &  3.30 &  3.30 & 3.30\\
 $6_{g}^{+}$&   6.70 & 6.70 & 6.80 &  6.82 &  6.83 & 6.84\\
 $8_{g}^{+}$&  11.14 &11.11 &11.42 & 11.47 & 11.50 &11.52\\
 $10_{g}^{+}$& 16.45 &16.34 &17.05 & 17.11 & 17.19 &17.23\\
 $12_{g}^{+}$& 22.47 &22.24 &23.57 & 23.62 & 23.79 &23.85\\
 $14_{g}^{+}$& 28.96 &28.68 &30.89 & 30.89 & 31.20 &31.27\\
 $16_{g}^{+}$& 35.60 &35.55 &38.91 & 38.81 & 39.32 &39.37\\
 $18_{g}^{+}$& 42.29 &42.78 &47.49 & 47.27 & 48.08 &48.07\\
 $20_{g}^{+}$& 49.30 &50.28 &56.75 & 56.20 & 57.39 &57.28\\
 $22_{g}^{+}$& 56.87 &58.01 &66.35 & 65.53 &       &66.92\\
 $24_{g}^{+}$& 65.07 &65.93 &76.28 & 75.19 &       &76.94\\
 $26_{g}^{+}$& 73.89 &74.00 &      & 85.14 &       &87.28\\
 $28_{g}^{+}$& 83.31 &82.20 &      & 95.34 &       &97.90\\
\noalign{\smallskip}\hline\noalign{\smallskip}
 $0_{\beta}^{+}$& 14.75 & 15.85  &  20.66  & 21.21 &22.56&22.51\\
 $2_{\beta}^{+}$& 15.55 & 16.85  &  21.43  & 22.21 &(23.38)&23.51\\
 $4_{\beta}^{+}$& 17.54 & 19.13  &  23.39  & 24.51 &     &25.81\\
 $6_{\beta}^{+}$&       & 22.55  &         & 28.03 &     &29.35\\
 $8_{\beta}^{+}$&       & 26.96  &         & 32.68 &     &34.03\\
\noalign{\smallskip}\hline
\end{tabular}
\end{center}
\vspace{-0.5cm}
\end{table}
\begin{table}[ht!]
\caption{Continuation of table \ref{tab:2}}
\label{tab:2c}
\begin{center}
\begin{tabular}{ccccccc}
\hline\noalign{\smallskip}
&\multicolumn{2}{c}{$^{160}$Dy}&\multicolumn{2}{c}{$^{162}$Dy}&\multicolumn{2}{c}{$^{164}$Dy}\\
\noalign{\smallskip}\hline\noalign{\smallskip}
$L$&Exp.&Th.&Exp.&Th.&Exp.&Th.\\
\noalign{\smallskip}\hline\noalign{\smallskip}
 $2_{\gamma}^{+}$& 11.13& 12.10  &11.01& 11.00 &10.38&10.23\\
 $3_{\gamma}^{+}$& 12.09& 12.94  &11.94& 11.91 &11.28&11.16\\
 $4_{\gamma}^{+}$& 13.32& 14.05  &13.15& 13.11 &12.48&12.38\\
 $5_{\gamma}^{+}$& 14.85& 15.41  &14.66& 14.60 &13.96&13.90\\
 $6_{\gamma}^{+}$& 16.58& 17.01  &16.42& 16.37 &15.75&15.69\\
 $7_{\gamma}^{+}$& 18.63& 18.83  &18.48& 18.40 &17.75&17.76\\
 $8_{\gamma}^{+}$& 20.74& 20.87  &20.71& 20.68 &20.04&20.09\\
 $9_{\gamma}^{+}$& 23.30& 23.11  &23.28& 23.20 &22.55&22.67\\
 $10_{\gamma}^{+}$&25.60& 25.52  &25.88& 25.94 &25.33&25.48\\
 $11_{\gamma}^{+}$&28.64& 28.11  &28.98& 28.90 &     &28.51\\
 $12_{\gamma}^{+}$&31.20& 30.85  &32.52& 32.06 &31.53&31.75\\
 $13_{\gamma}^{+}$&34.44& 33.72  &35.45& 35.40 &     &35.19\\
 $14_{\gamma}^{+}$&37.11& 36.73  &39.00& 38.92 &     &38.81\\
 $15_{\gamma}^{+}$&40.43& 39.85  &42.57& 42.60 &     &42.61\\
 $16_{\gamma}^{+}$&43.42& 43.08  &46.30& 46.44 &     &46.56\\
 $17_{\gamma}^{+}$&46.60& 46.40  &50.08& 50.41 &     &50.67\\
 $18_{\gamma}^{+}$&50.13& 49.82  &53.84& 54.51 &     &54.92\\
 $19_{\gamma}^{+}$&53.22& 53.31  &     & 58.74 &     &59.29\\
 $20_{\gamma}^{+}$&57.33& 56.88  &     & 63.08 &     &63.79\\
 $21_{\gamma}^{+}$&60.39& 60.51  &     & 67.52 &     &68.41\\
 $22_{\gamma}^{+}$&64.55& 64.21  &     & 72.05 &     &73.12\\
 $23_{\gamma}^{+}$&68.18& 67.96  &     & 76.68 &     &77.94\\
 $24_{\gamma}^{+}$&     & 71.76  &     & 81.39 &     &82.84\\
 $25_{\gamma}^{+}$&76.54& 75.61  &     & 86.18 &     &87.83\\
\noalign{\smallskip}\hline\noalign{\smallskip}
$\chi$&\multicolumn{2}{c}{0.423} &\multicolumn{2}{c}{0.067} &\multicolumn{2}{c}{0.734}\\
$a$  &\multicolumn{2}{c}{14.519}&\multicolumn{2}{c}{7.934}&\multicolumn{2}{c}{24.473}\\
$\beta_{0}$  &\multicolumn{2}{c}{2.442}&\multicolumn{2}{c}{3.056}&\multicolumn{2}{c}{3.205}\\
$\sigma$  &\multicolumn{2}{c}{0.636}&\multicolumn{2}{c}{0.411}&\multicolumn{2}{c}{0.092}\\
\noalign{\smallskip}\hline
\end{tabular}
\end{center}
\vspace{-0.5cm}
\end{table}

The normalized theoretical and experimental energies are listed in Table \ref{tab:1} for Gd isotopes and in Tables \ref{tab:2} and \ref{tab:2c} for those of Dy, where the resulting $\sigma$ measure and the fitted values of the three parameters are also presented. As can be seen, the agreement with experiment for all treated nuclei is very good considering the number of fitted energy states. Especially well are reproduced the ground and $\gamma$ band energies. The energy spectra of Dy isotopes are overall better described even though the fits were performed on more experimental states: 14, 18, and 12 for $^{158}$Gd, $^{160}$Gd and $^{162}$Gd against 39, 31, and 20 for $^{160}$Dy, $^{162}$Dy and $^{164}$Dy. This is also true when comparing with the results of other models with the same number of free parameters \cite{BonMD1,BonMD2} based on modified inertial parameters. Indeed, in \cite{BonMD1} and \cite{BonMD2} where a Davidson and respectively a Kratzer potential was used in connection to a deformation dependent mass term, the energy spectrum of Gd nuclei where better treated than in the present calculations, while the Dy ones poorer. However, one must mention here that for the middle Gd and Dy nuclei a different $\beta$ experimental band was taken into account for comparison as per suggestions of \cite{160GdDy,162GdDy}.

\begin{figure*}[th!]
\begin{center}
\includegraphics[clip,trim = 0mm 0mm 0mm 0mm,width=0.6\textwidth]{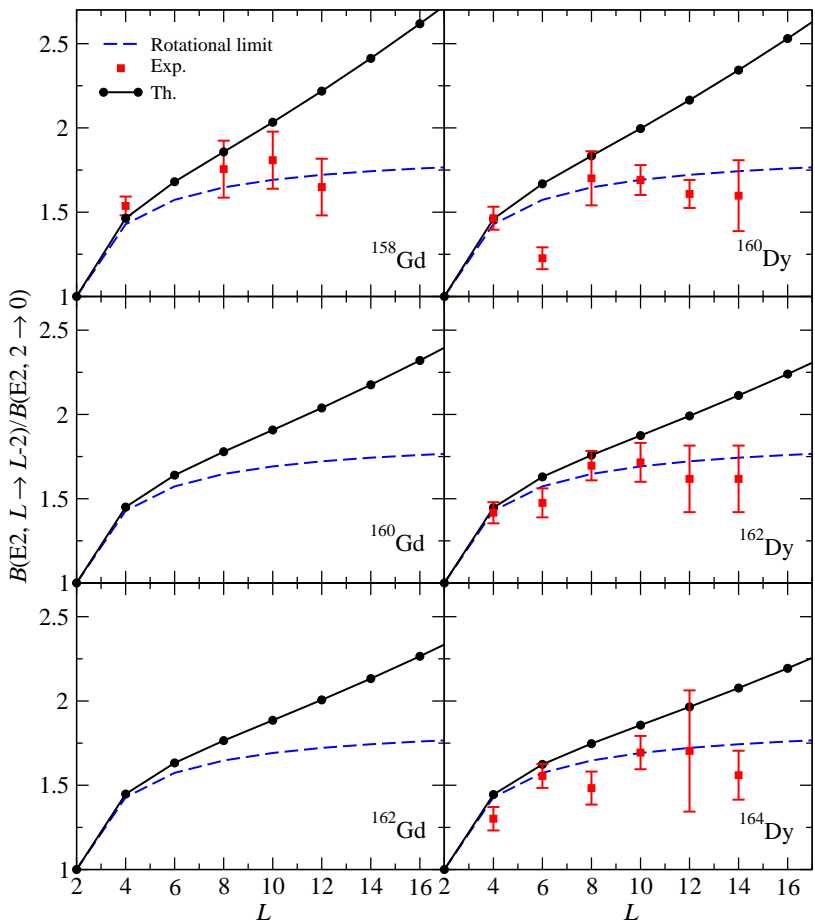}
\end{center}
\vspace{-0.2cm}
\caption{Theoretical ground state to ground state $E2$ transition probabilities normalized to the $2_{g}^{+}\rightarrow0^{+}_{g}$ transition are compared with the available experimental data corresponding to all considered nuclei and with the rigid rotor predictions.}
\label{tg}
\end{figure*}

The resulting set of parameters from the fitting procedure are further used to compute the ratios for some relevant $B(E2)$ rates. Note that although here one uses the $\chi$ dependent integration measure instead of the usual one used in Ref.\cite{Noi}, the correction have an almost negligible effect on the numerical values for most of the transitions. The exception are the $\beta$-ground transition rates, where the effect is more distinguishable and which however were not considered in Ref.\cite{Noi} due to the lack of experimental data for the nuclei considered there. The theoretical results for the ground to ground transitions normalized to the $B(E2,2_{g}^{+}\rightarrow\,0_{g}^{+})$ value are shown graphically in Fig.\ref{tg} together with the available experimental data points, while those corresponding to low lying interband transitions are compared with scarce experimental values in Table \ref{tab:3}. The later ones are normalized to the lowest transition of the same kind in order to ascertain the specific relationship between the interacting bands. As the considered nuclei are generally accepted as being strongly deformed, the rigid rotor assumption inevitably comes into discussion. Although the comparison of energy states is obviously not relevant due to great differences (deviations from rotational formula, nondegeneracy of the $\beta$ and $\gamma$ bands), the rigid rotor predictions serve as a useful reference for the transition probabilities. Such that, one included in Fig.\ref{tg} and Table \ref{tab:3} also the rigid rotor limit. With this one can see that the experimental points from Fig.\ref{tg} are situated generally in between the rigid rotor and present model's predictions. Moreover, the data are closer to the rigid rotor limit in case of the Dy isotopes, while the only measurements for the Gd isotopes associated to $^{158}$Gd are closer to the present calculations. The same happens for the interband transitions listed in Table \ref{tab:3}, where the agreement with experiment for $^{158}$Gd nucleus is even better. Due to the lack of experimental data for $\beta$-ground transitions in the remaining nuclei one can only point that the present predictions are greater than those corresponding to the rigid rotor limit. In case of the $\gamma$-ground transitions the calculations offer slightly higher values than the rigid rotor formula and both of them underestimate the data. The agreement with experiment for the interband transitions is similar to the calculations of Refs.\cite{BonMD1,BonMD2}, while for the ground-ground transition the agreement is better especially for higher angular momentum states.

\setlength{\tabcolsep}{6.5pt}
\begin{table}[th!]
\caption{Theoretical estimations of the ratios corresponding to few low lying interband $B(E2)$ transition probabilities for all considered nuclei are compared with available experimental data. $\Delta K=0$ transitions are normalized to the $2_{\beta}^{+}\rightarrow0^{+}_{g}$ transition, while $\Delta K=2$ transitions to the $2_{\gamma}^{+}\rightarrow0^{+}_{g}$ transition, as in Ref.\cite{BonDES,Bijker}. R.R. stands for Rigid Rotor.}
\label{tab:3}
\begin{center}
\begin{tabular}{cllll}
\hline\noalign{\smallskip}
Nucleus& $\frac{2^{+}_{\beta}\rightarrow2^{+}_{g}}{2^{+}_{\beta}\rightarrow0^{+}_{g}}$& $\frac{2^{+}_{\beta}\rightarrow4^{+}_{g}}{2^{+}_{\beta}\rightarrow0^{+}_{g}}$& $\frac{2^{+}_{\gamma}\rightarrow2^{+}_{g}}{2^{+}_{\gamma}\rightarrow0^{+}_{g}}$& $\frac{2^{+}_{\gamma}\rightarrow4^{+}_{g}}{2^{+}_{\gamma}\rightarrow0^{+}_{g}}$\\
\noalign{\smallskip}\hline\noalign{\smallskip}
$^{158}$Gd&0.25(6)&4.48(75)&1.76(26)&0.079(14)\\
&1.93&6.01&1.46&0.077\\
$^{160}$Gd&&&1.87(12)&0.189(29)\\
&1.79&4.97&1.44&0.074\\
$^{162}$Gd&&&&\\
&1.76&4.80&1.44&0.074\\
$^{160}$Dy&&2.52(44)&1.89(18)&0.133(14)\\
&1.89&5.70&1.45&0.075\\
$^{162}$Dy&&&1.78(16)&0.137(12)\\
&1.75&4.70&1.44&0.073\\
$^{164}$Dy&&&2.00(27)&0.240(33)\\
&1.73&4.55&1.44&0.073\\
R.R.&1.43&2.57&1.43&0.071\\
\noalign{\smallskip}\hline
\end{tabular}
\end{center}
\end{table}

Despite having an additional free parameter comparing to the approach of Ref.\cite{Noi}, namely $\beta_{0}$, the present model gives a higher $\sigma$ value for the same $^{160}$Gd nucleus. However both results are consistent in what concerns the resulting values for the rigidity and stiffness parameters, i.e. $\chi$ and $a$. Indeed, in both calculations, a high $\gamma$-rigidity is predicted corroborated also with a large stiffness of the $\gamma$ oscillation, which supports the statement made in Section 2 regarding the interdependence between the rigidity and stiffness parameters. The numerical application for its neighbours reveals a quite distinct character of these nuclei expressed through a very small $\chi$ value corresponding to a high $\gamma$-stability. The quasi $\gamma$-rigid nature of the $^{160}$Gd nucleus is also confirmed by the relativistic mean-field calculations \cite{Niksic} of the quadrupole binding energy $\beta-\gamma$ maps of the Gd nuclei up to $^{160}$Gd, which present the deepest and most localized minimum in respect to the $\gamma$ shape variable. The singularity behaviour of the $^{160}$Gd isotope in regard to the $\gamma$-rigidity of its neighbours is also found in the Dy isotopes but less strikingly and with an opposite effect. While the $^{162}$Dy nucleus is near $\gamma$-stable, its neighbours exhibit a fair amount of $\gamma$-rigid/stable mixing. From the present model's point of view there is definitely a critical aspect in the behaviour of the $^{160}$Gd and $^{162}$Dy. This is supported by the fact that the $\gamma$-rigidity parameter $\chi$ which discontinuously evolve along the two sets of isotopes has the biggest effect on the energy spectrum as was established when Fig.\ref{spec} was analysed. The experimental traces of the turning point realized in these nuclei must be searched in the structure of their $\beta$ and $\gamma$ bands which are most affected by the rigid-stable coupling. The structure of the $\gamma$ band is often studied with the help of the quantity $S(4)$ which is defined as \cite{Zamfir}:
\begin{equation}
S(L)=\frac{\left[E(L_{\gamma})-E(L_{\gamma}-1)\right]-\left[E(L_{\gamma}-1)-E(L_{\gamma}-2)\right]}{E(2^{+}_{g})}
\end{equation}
and measures the $\gamma$ band staggering. A similar measure, but aimed to describe the relative spacing of the lowest states in the $\beta$ band is given by \cite{BonDES}:

\begin{figure}[th!]
\begin{center}
\includegraphics[clip,trim = 0mm 0mm 0mm 0mm,width=0.49\textwidth]{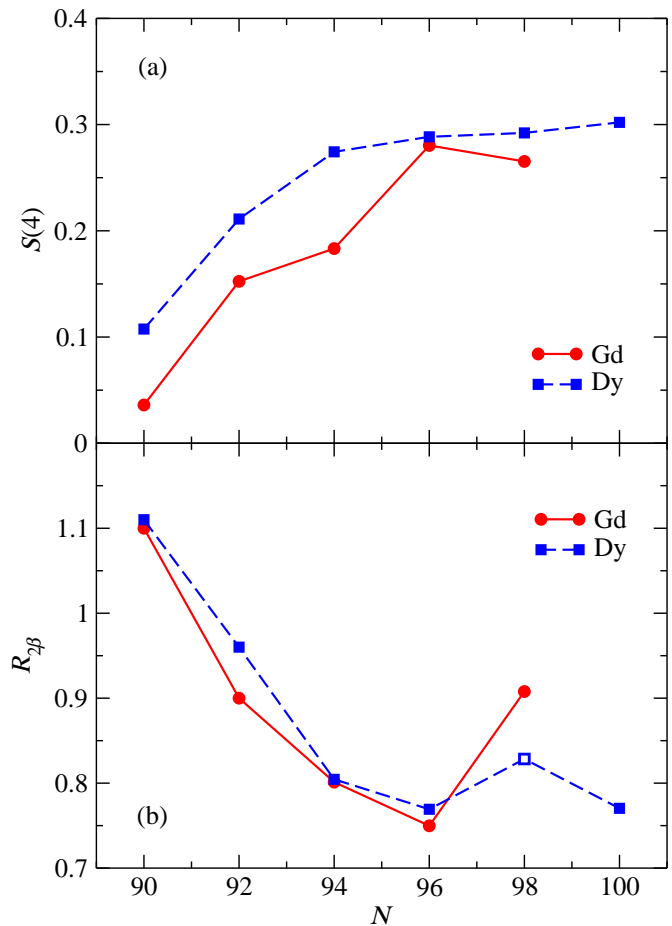}
\end{center}
\vspace{-0.2cm}
\caption{$S(4)$ (a) and $R_{2\beta}$ (b) quantities as function of the neutron number $N$ in case of the Gd and Dy isotopic chains, calculated using the experimental energies from Refs.\cite{158Gd,160GdDy,162GdDy,164Dy,166,168}. The open symbol corresponds to a value calculated using uncertain experimental data.}
\label{exps}
\end{figure}

\begin{equation}
 R_{2\beta}=\frac{E(2_{\beta}^{+})-E(0_{\beta}^{+})}{E(2_{g}^{+})}.
\end{equation}
The experimental values for both of these quantities are plotted in Fig.\ref{exps} as function of the neutron number $N$ in case of the Gd and Dy isotopic chains. From the Fig.\ref{exps}(a) one can see that although the $\gamma$ staggering steadily increases with the number of neutrons, a local peak is registered at $N=96$ which is sharper for the Gd chain in comparison to the Dy isotopes where it is barely visible. The critical behaviour of the $N=96$ isotopes is more obviously reflected in the plot of the $R_{2\beta}$ signature in Fig.\ref{exps}(b). Indeed, the function $R_{2\beta}=f(N)$ has clear minima at $N=96$ for both isotopic chains, and once again the minimum is sharper for the Gd chain. The more defined criticality of the $^{160}$Gd nucleus is due to the bigger differences obtained in the values of the parameter $\chi$ associated to its neighboring nuclei. However, the singularity behaviour of $^{160}$Gd and $^{162}$Dy nuclei shown in Fig.\ref{exps} do not hint to the opposite effect found in relation to the $\gamma$-rigidity parameter values. An experimental validation of this distinction between the two isotopic chains might come from data on the $E2$ transitions connecting low states of the $\gamma$ or $\beta$ bands. Indeed, although the picture is not complete, one can see in Table \ref{tab:3} that the experimental values of the $B(E2,2_{\gamma}^{+}\,\rightarrow\,2_{g}^{+})/B(E2,2_{\gamma}^{+}\,\rightarrow\,0_{g}^{+})$ ratio for the $^{160}$Dy nucleus is smaller than that of its neighboring isotopes, while the same ratio corresponding to $^{160}$Gd have a greater value than its lighter isotope.

With the above analysis it is clear that in the neutron rich part of both Gd and Dy isotopic chains a critical phenomenon takes place. Even if its nature was shown to be related to the interplay between the $\gamma$-rigid and $\gamma$-stable excitations it must be further investigated.

\section{Conclusions}
Starting form the classical picture of a mixed $\gamma$-rigid/stable nucleus in the framework of LDM, the theoretical foundations of the model proposed in ref.\cite{Noi} where established on solid grounds removing in the same time some of its drawbacks. Moreover, adopting a Davidson potential instead of the infinite square well used in Ref.\cite{Noi}, one obtained a more flexible formalism in respect to the reproduction of the experimental data but with the price of an additional parameter. As a result the analytical formulas for the energy spectrum depend on three parameters, namely the $\gamma$-rigidity $\chi$ which mediates the interplay between the two types of $\beta$ vibration, the stiffness of the $\gamma$ stable vibration and $\beta_{0}$ - the minimum position of the Davidson potential. The effect provided by the first two parameters is basically the same as in Ref.\cite{Noi}. In what concerns the additional parameter $\beta_{0}$, it was found that its variation increases very rapidly the ground and $\beta$ band states softening also their $\chi$-dependence. This overriding of the $\gamma$-rigidity managed by $\chi$ is due to the precedence of the $\beta$ shape variable over the $\gamma$ one.  Another interesting finding is the fact that the $\gamma$ band head responds quite weakly to the $\beta_{0}$ variation.

The model was successfully applied in case of the $^{160}$Gd and $^{162}$Dy nuclei previously described within the formalism of Ref.\cite{Noi}. Moreover, due to the more pliable Davidson potential, the model also provided a very good quantitative description of the energy spectra and electromagnetic properties for their neighbouring isotopes, i.e. $^{158}$Gd, $^{162}$Gd, $^{160}$Dy and $^{164}$Dy. An impressive agreement with experimental energy spectra is obtained in case of the Dy isotopes, while the available experimental $E2$ transition probabilities are better reproduced for the $^{158}$Gd nucleus.

The values of the fitted parameters for $^{160}$Gd and $^{162}$Dy are consistent with those obtained in Ref.\cite{Noi}. However the values obtained for $\chi$ corresponding to their neighboring isotopes are quite dissimilar. Indeed, the nucleus $^{160}$Gd is predominantly $\gamma$-rigid while its neighbors clearly prefer the $\gamma$-stable conditions. The picture is reversed in case of the Dy isotopes and with less radical differences. As the major influence on the present model is exercised by the $\gamma$-rigidity parameter $\chi$, this result is not just a numerical peculiarity. As a matter of fact, studying some relevant spectral signatures one identified the same singularity character of the $^{160}$Gd and $^{162}$Dy nuclei in respect to their isotopic chains. In what concerns the opposite $\gamma$-rigidity feature of Dy and Gd isotopes, traces of it are identified in the experimental $\gamma$-ground $E2$ transition probabilities.

The occurrence of $^{160}$Gd and $^{162}$Dy as singular points in their respective isotopic chains due to abnormal collective behaviour, to our knowledge was never discussed before from the theoretical point of view. Such that the fact that in the framework of the present model, the phenomenon is attributed to sudden and sizeable change in the ratio between the $\gamma$-rigid and $\gamma$-stable apport to the collective motion, represents an important step in the understanding of the evolution of collectivity in the neutron rich transitional nuclei.

The present and previous \cite{Noi} results of the newly established approach based on the competition between the $\gamma$-rigid and $\gamma$-stable excitations encourage us to attempt a similar treatment for triaxial nuclei. Such a project would be able to make a bridge between the $Z(4)$ \cite{BonZ4} and $Z(5)$ \cite{BonZ5} related models \cite{Fortunato2,Fortunato3,Z5D}, bringing new insights into the dynamics of triaxial nuclei.

%
%

\end{document}